\documentclass[aps,pre,twocolumn,showpacs,groupedaddress,showkeys,amsmath,amssymb]{revtex4}
\usepackage{color}
\usepackage{epsfig}
\usepackage{graphicx}
\begin{document}
\title{Study of the $^4$He crystal surface}
\author{K.O.\,Keshishev, V.I.\,Marchenko, and D.B.\,Shemyatikhin\\
Kapitza Institute for Physical Problems, RAS, Moscow, 119334
Russia}
\date{\today}
\begin{abstract}The evolution of the meniscus of a helium crystal near
the (0001) face is traced during a change in the boundary
conditions at the chamber wall in the temperature range 0.5-0.9 K.
The critical behavior of the contact angle is studied. An
anisotropy is detected in the crystal-glass interface energy. New
data on the temperature dependence of the elementary-step energy
are obtained.
\end{abstract}
\maketitle

In this work, we continue to study the behavior of the contact
angle that appears when the interface of two condensed phases of
$^4$He (crystal-superfluid) reaches a solid wall. The formulation
of the problem and a detailed description of our optical technique
were given in our earlier works \cite{KSS,KS}. Recall only that
the technique consists in photographing a crystal using parallel
light followed by digital processing of images. To improve the
images, we substantially modified the cryostat design. In the
modified version, an optical tract passes through windows located
in the vacuum space of the cryostat and does not meet liquid
helium. As a result, we were able to decrease the image noise by
several times.

\begin{figure}[b!]\center
\includegraphics[width=0.7\columnwidth]{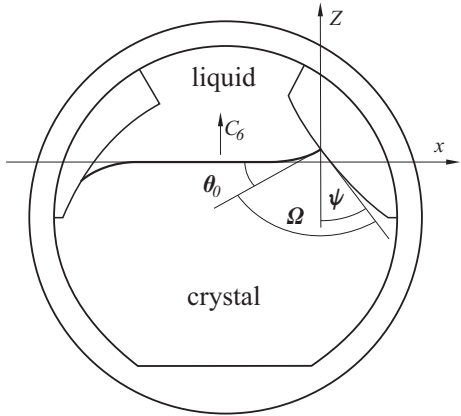}
\caption{Cross section of the experimental chamber}
\label{camera}\end{figure}

Figure\,\ref{camera} schematically shows the cross section of the
experimental chamber with a crystal at the bottom. Recall also
that, when analyzing the meniscus profile near the chamber walls,
we consider a two-dimensional problem in the section plane, since
the longitudinal chamber size (29\,mm) is significantly larger
than the capillary constant ${\lambda\sim1}$\,mm. The crystal is
oriented so that the (0001) basal plane is horizontal. The $x,z$
plane of a Cartesian coordinate system coincides with the section
plane and axis $z$ is vertical. Due to the chosen configuration of
the side walls, the interface has an $S$-like shape described by a
$Z(x)$ function. Then, angle ${\theta=arctg(dZ/dx)}$ determines
the surface inclination at a given point with respect to the
(0001) plane. At the point of the end of surface at the wall, we
have ${\theta=\theta_0},$ and $\theta_0$ is one of the quantities
to be measured in our experiment. The second quantity to be
measured is the angle of wall inclination $\psi$ to axis $C_6$ of
the crystal at the contact point.

Angle of contact with the right wall $\Omega_R$ is
\begin{equation}\label{R}\Omega_R=\psi-\theta_{R0}+\frac{\pi}{2}.\end{equation}
It is related to the surface energy of the crystal
$\alpha(\theta),$ crystal-wall energy $\varepsilon_s,$ and
liquid-wall energy $\varepsilon_l$ by the equation
\begin{equation}\label{RR}
\alpha(\theta_{R0})\cos\Omega_R+\alpha'_\theta(\theta_{R0})\sin\Omega_R=
-\Delta\varepsilon,\end{equation} where
${\Delta\varepsilon=\varepsilon_s-\varepsilon_l}$

The angle of contact with the left wall is
\begin{equation}\label{L}\Omega_L=\psi+\theta_{L0}+\frac{\pi}{2}.
\end{equation}

The boundary condition near the left wall has the form
\begin{equation}\label{RL}
\alpha(\theta_{L0})\cos\Omega_L-\alpha'_\theta(\theta_{L0})\sin\Omega_L=
-\Delta\varepsilon.\end{equation}

The right and left walls are made of the same material (polished
glass). Therefore, the same physical state should take place at
both walls at the same values of $\psi$ for the horizontal
orientation of the basal plane in the crystal due to the presence
of screw twofold axis $C_2$ in the symmetry group of the crystal.
When $C_2$ rotates, angle $\theta_{L0}$  changes its sign and we
will present experimental data for the general function
${\theta_0(\psi)=\theta_{R0}(\psi)=-\theta_{L0}(\psi)}.$

\begin{figure}[t!]\center
\includegraphics[width=0.8\columnwidth]{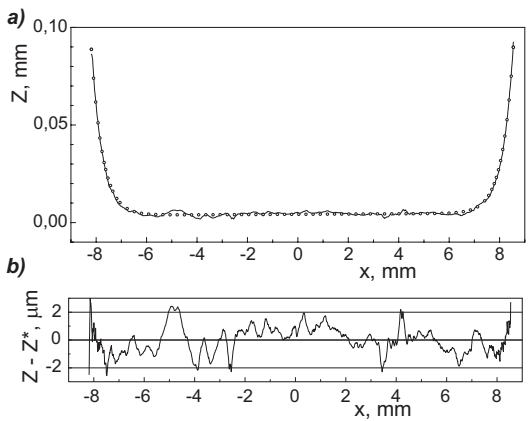}
\caption{\textit{a)} Shape $Z(x)$ (line) of the meniscus of liquid
$^4$He and (circles) calculated profile. \textit{b)} Deviation of
the experimental from the calculated curve.} \label{liquid}
\end{figure}

To control the possibilities of our optical technique, we used the
results of digital processing of photographs taken from two test
samples. Figure\,\ref{liquid}à shows shape $Z(x)$ of the meniscus
of liquid\,$^4$He filling the lower part of the chamber. Here, we
also present the surface shape calculated in the low-angle
approximation ${\theta\ll1},$
\[Z^*(x)=A+B\cosh\frac{x}{\Lambda},\,\Lambda=\sqrt{\frac{\alpha_l}{\rho_l g}},\] where
${\alpha_l=0{,}28}$\,erg/cm$^2$ ---  is the surface tension of
liquid\,$^4$He at\,${T=2{,}3}$\,K\,\cite{AM},
${\rho_l=0{,}145}$\,g/cm$^2$ --- is its density, and $g$ --- is
the gravitational acceleration. Figure\,\ref{liquid}b shows the
dependence ${{\delta}Z(x)=Z(x)-Z^*(x),}$ which characterizes the
deviation of an experimental curve from the calculated curve
corresponding to the optimum choice of adjustable parameters $A$
and\,$B.$

\begin{figure}[t!]\center
\includegraphics[width=0.8\columnwidth]{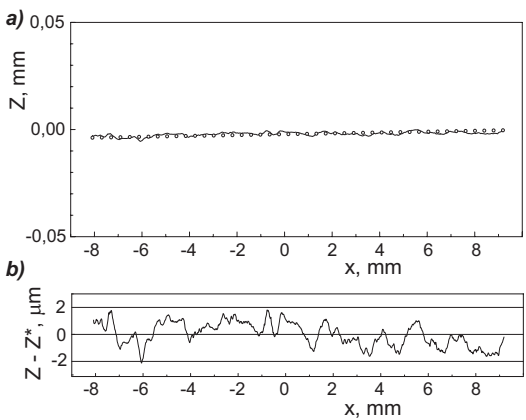}
\caption{\textit{a)} Shape $Z(x)$ of a flat growing face of a
$^4$He crystal, and the dashed line shows the face direction.
\textit{b)} Deviation of the experimental curve from a straight
line.} \label{facet}
\end{figure}

Similar data are presented in Fig.\,\ref{facet} for the crystal
the (0001) plane of which is close to the horizontal. This
photograph was taken at a temperature of\,0.88\,K, i.e., well
below roughening temperature\,$T_R$ of the basal plane. Under
these conditions, the interface coincides with the growing face.
In Fig.\,\ref{facet}a, dependence $Z(x)$ is approximated by a
straight line ${Z^*=A+Bx}.$ The difference between the
experimental results and the optimum straight line is shown in
Fig.\,\ref{facet}b. In both cases
(Figs.\,\ref{liquid}b,\,\ref{facet}b), the deviations from the
real shape are random (about ${\pm2\,\mu}$m) and are the main
origin of errors in determining the surface shape. The image of
the liquid helium meniscus can be used to detect the horizontal
direction accurate to\,${\sim 2\cdot10^{-4}}$\,rad. Using
photographs of a growing face, its direction in the $xz$ plane was
determined accurate to\,${\sim 2\cdot10^{-4}}$\,rad.

At the initial stage of experiments, we grew a small
${\sim1}$\,mm$^3$ crystal whose basal plane was close to the
horizontal. This procedure was described in detail in
\cite{KSS,KS}. The crystal was then grown slowly to reach the
level of the glass walls. The growth rate did not exceed
${\sim1\,\mu}$m/s and was stabilized by controlled heating of the
ballast volume outside the cryostat. The pressure in the ballast
volume exceeded the equilibrium pressure by\,${\sim0{.}4}$\,mbar.

After the chosen temperature was stabilized, we took photographs
of the crystal profiles at various levels. The transitions to the
next levels were performed by sequential surface melting or
growing of the crystal. The difference between neighboring levels
was\,0.2--0.5\,mm.

Once the next level was reached, the chamber with the crystal was
disconnected from the ballast volume with a valve located outside
the cryostat at room temperature (unfortunately, the design of our
device does not imply the existence of a "cold"\, valve). As a
result, quasi-equilibrium conditions, which are accompanied by a
very slow helium flow from the filling line to the chamber because
of a low level of liquid helium in the cryostat bath at\,4.2\,K,
were established. Under these conditions, the crystal grew slowly
at a rate of ${\sim5\cdot10^{-3}\,\mu}$m/s and had almost the same
shape.

\begin{figure}[b!]\center
\includegraphics[width=0.8\columnwidth]{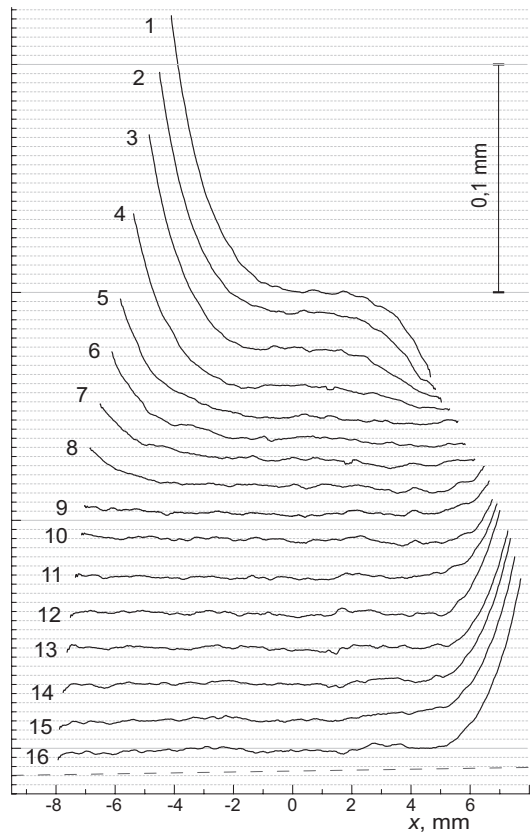}
\caption{Series of crystal menisci at ${T=0{,}61}$\,K. The face
direction is indicated by the heavy dashed line. The photographs
were taken during melting.} \label{061}
\end{figure}

\begin{figure}[b!]\center
\includegraphics[width=0.8\columnwidth]{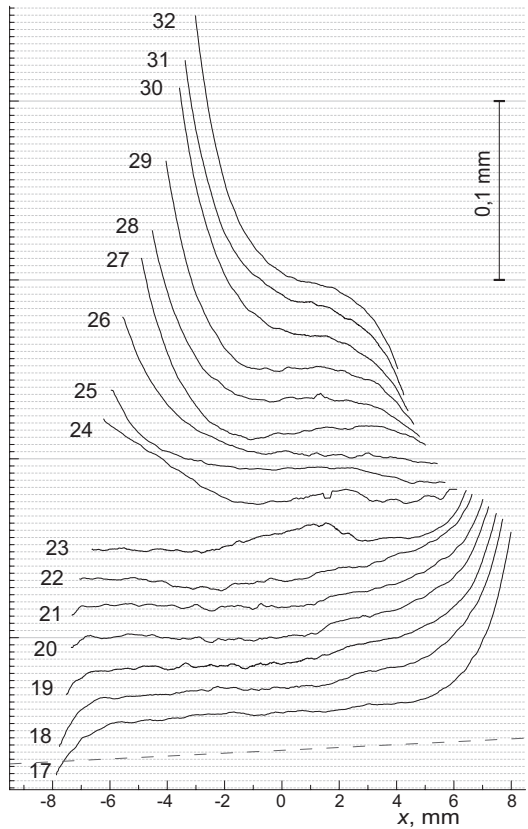}
\caption{Series of crystal menisci at ${T=0{,}9}$\,K. The face
direction is indicated by the heavy dashed line. The photographs
were taken during growth.} \label{09}
\end{figure}

We now present the measurement results for two crystals. In the
first sample, the basal plane was inclined at an angle of
${2\cdot10^{-4}}$\,rad in the transverse direction (with respect
to axis $x$) and at an angle of ${1,2\cdot10^{-3}}$\,rad  in  the
longitudinal direction (with respect to axis y). For the second
sample, the transverse inclination was ${8\cdot10^{-4}}$\,rad and
the longitudinal inclination was ${6\cdot10^{-4}}$\,rad. After
measuring the longitudinal face inclination, the optical bench was
inclined at the measured angle; as a result, the error in
parallelism between the optical axis and the crystal face was
corrected.

The first sample was photographed during gradual surface melting,
and the second sample was grown gradually. In both cases,
photographs were taken in 20 and\,40\,min after the chamber was
closed with the valve in order to control surface relaxation. For
the first sample, we took two series of photographs corresponding
to temperatures of 0.89 and\,0.61\,K; for the second sample, three
series of photographs were taken at temperatures of 0.9,\,0.72,
and\,0.53\,K. The results of processing two of the five series are
shown in Fig.\,\ref{061} (first sample, ${T=0{,}61}$\,K) and
Fig.\,\ref{09} (second sample, ${T=0{,}9}$\,K). The scale in the
ordinate axis is higher than that of the abscissa scale by more
than\,60\,times. All curves are located as close as possible to
each other without conserving the vertical scale. The real
vertical distance between the centers of the upper and lower
profiles is about\,5\,mm.

The situation seems to be ambiguous from the standpoint of an
equilibrium surface shape.

An equilibrium surface shape is known to correspond to the minimum
of the surface and gravitational energy, and the surface rigidity
plays a key role in this case. In our case, we are dealing with
so-called longitudinal rigidity
${\tilde{\alpha}=\alpha+\alpha''}.$ The surface rigidity was
studied by various methods\,\cite{RGBRN,BKP,AKO}.  The result
substantial for us consist in the following. In the temperature
range ${0{.}4\,\mathrm{K}<T<T_R},$ the surface rigidity has a weak
anisotropy for atomically rough surface regions at sufficiently
low angles of inclination (${\theta\lesssim0{,}1}$\,rad). The
rigidity is temperature-independent
(${\tilde{\alpha}_0\approx0{,}245}$\,erg/cm$^2$). A sharp decrease
in \,$\tilde{\alpha}$ was only detected \,\cite{RGCB} at
temperatures below 0{,}3\,K and low angles ${\theta<0{,}04}$\,rad.

\begin{figure}[h]\center
\includegraphics[width=0.7\columnwidth]{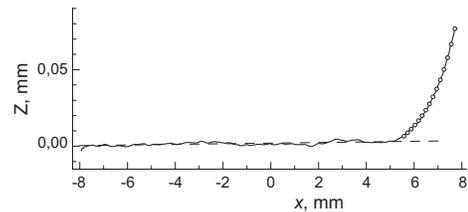}
\caption{Profile 16 (line) at ${T=0{,}61}$\,K and (circles)
approximation of the meniscus.} \label{N061}
\end{figure}

For all series of our experimental data, surface regions at not
very low angles of inclination
(${10^{-2}\lesssim\theta\lesssim0{,}1}$\,rad) obey the equilibrium
equation ${\tilde{\alpha}Z''_{xx}=\Delta{\rho}Z,}$ where
${\tilde{\alpha}=0{,}245}$\,erg/cm$^2,$ and $\Delta\rho$ is the
difference between the densities of the solid and liquid phases.
Fig.\,6 shows the approximation of the profile 16 by the solution
to Eq.\,(\ref{z}), ${Z=A\sinh\frac{x-x_0}{\Lambda}},$ where
${\Lambda=(\tilde{\alpha}/\Delta{\rho}g)^{1/2}}.$ The dashed line
shows the measured direction of the (0001) face inclined at
$2\cdot10^{-4}$\,rad to the horizontal.

Note that the meniscus shape can also be described \cite{KS} with
a satisfactory accuracy by the function ${Z=\pm(x-x_0)^3b^{-2}},$
where ${b\sim1}$\,mm, which corresponds to the standard theory of
a vicinal surface as the echelon of steps that repel each other as
$\propto x^{-2}.$ However, this picture is in conflict with the
results of measuring $\tilde{\alpha}$ from the spectrum of
crystallization waves at not very low angles.

\begin{figure}[h]\center
\includegraphics[width=0.7\columnwidth]{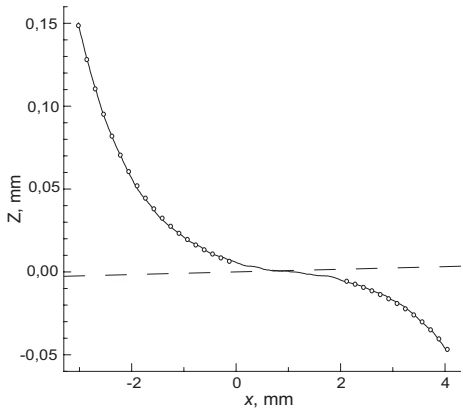}
\caption{ Profile 32 (line) at ${T=0{,}9}$\,K and its
approximation (circles).} \label{N09}
\end{figure}

Similar results of processing profile 32 Fig.\,\ref{09} are shown
in\,Fig.\ref{N09}. In this case, the left and right segments of
the profile are approximated by the formula
${Z_\pm=A\sinh\frac{x-x_\pm}{\Lambda}}$ at the same amplitude $A$
and parameters $x_\pm,$ differing by\,$\approx1$\,mm.

Thus, we can state that the surface is in equilibrium at not very
low angles ${10^{-2}\lesssim\theta\lesssim10^{-1}}.$ The situation
is  radically different  at  angles
${\theta\lesssim10^{-2}}$\,rad: non-monotonic profiles often
appear, which indicates the absence of equilibrium (see
Figs.\,\ref{09},\ref{061}). The lengths of such regions along axis
$x$ change within several millimeters, and the irregular
deviations from the vertical are ${10--20\,\mu}$m. Such phenomena
are usually related to lattice defects, mainly dislocations, and
we cannot exclude this explanation. We only recall that the
crystals were grown with all precautions necessary in such cases.

However, the shapes of some surface fragments can hardly be
explained by the presence of dislocations. For example, the center
portion of profile 23 (Fig.\,\ref{09}) contains a linear (within
the limits of experimental error) ${\sim4}$\,mm segment inclined
at an angle of 0.003\,rad to the (0001) face. The next frame
(profile 24) contains a flat region reaching the left wall and
inclined at an angle of 0{.}01\,rad to the opposite side. The
presence of such extended surface regions is thought to be hardly
explained by the existence of defects in the crystal volume.
Unfortunately, we have no certain considerations regarding the
nature of these metastable states. It is difficult to draw any
quantitative conclusions concerning the equilibrium meniscus shape
(and the contact angle) and, hence, the angular dependence of the
surface rigidity at low angles under these conditions.

\begin{figure}[t]\center
\includegraphics[width=0.8\columnwidth]{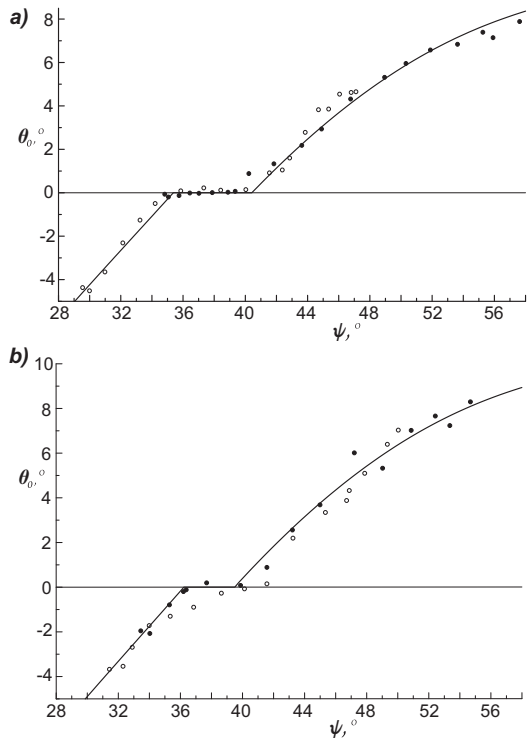}
\caption{Dependence $\theta_0(\psi)$ for two series of profiles
obtained at \textit{a)} ${T=0{,}61}$\,K  è \textit{b)}
${T=0{,}9}$\,K. Open and solid symbols correspond to the right and
left sides of a meniscus, respectively. (solid line) Calculation
by Eqs.\,(\ref{RR}),(\ref{RL}),(\ref{alpha}), and (\ref{De}).}
\label{theta_psi}
\end{figure}

The authors of \cite{MP} theoretically predicted a phenomenon
caused by the jump of derivative $\alpha'_\theta$  at
${\theta=0}.$ This phenomenon consists in the fact that the state
where an atomically smooth face is in immediate contact with the
wall takes place in the angular range ${\psi_-<\psi<\psi_+},$
determined by the relation
\[|\Delta\varepsilon-\alpha_0\sin\psi|<\beta\cos\psi\]
where ${\beta=\alpha'_\theta>0}$ at ${\theta=+0}.$ In this case, a
plateau should appear in the $\theta_0(\psi)$ dependence in a
certain angular range. Figure\,\ref{theta_psi} shows the results
of processing our photographs for this dependence. The plateau is
seen to exist. However, the $\theta_0(\psi)$ dependence differs
substantially from the root behavior suggested in \cite{MP}. The
contact angle approaches the plateau linearly at not very low
angles ${\theta_0\sim1^\circ}.$

Recall that the meniscus profile was determined from an analysis
of a diffraction pattern, as was described in our earlier works
\cite{KSS,KS}. However, it is difficult to perform this analysis
near the crystal-liquid-wall contact line, i.e., in the region
where the contact angle is to be determined. Here, the diffraction
pattern is complicated because of the approaching of two
relatively simple diffraction patterns from the wall and the
crystal-liquid interface at a distance of ${\approx0{.}3}$\,mm,
which is comparable with ${\Lambda\approx1{.}2}$\,\,mm. The value
of $\theta_0$ at ${\theta_0>0{,}01}$\,rad under these conditions
was determined by the extrapolation of the derivative of the
function approximating the meniscus
${Z=A\sinh\frac{x-x_0}{\Lambda}}$ to the calculated point of
contact with the wall. The region of low contact angles
${\theta_0\leq0{.}01}$\,rad, where the root law is likely to be
valid, requires an additional investigation.

The observed behavior of the meniscus at angles
${10^{-2}<\theta\lesssim10^{-1}}$\,rad can be explained under the
following assumptions: function $\alpha(\theta)$ at these angles
has the form
\begin{equation}\label{alpha}\alpha=\alpha_0+
\beta|\theta|+\alpha''_0\frac{\theta^2}{2},\end{equation} and
solid helium-glass interface energy $\varepsilon_s$ is a function
of angle\,$\psi.$ Moreover, we believe that, in the temperature
range under study, only the step energy changes with  temperature
and the other parameters ${\alpha_0\approx0{,}172}$\,erg/cm$^2$
\cite{AK91},
${\alpha''_0=\tilde{\alpha}_0-\alpha_0\approx0{,}073}$\,erg/cm$^2,$
and function $\Delta\varepsilon(\psi)$ almost reached their values
characteristic of zero temperature.

\begin{figure}[h]\center
\includegraphics[width=0.75\columnwidth]{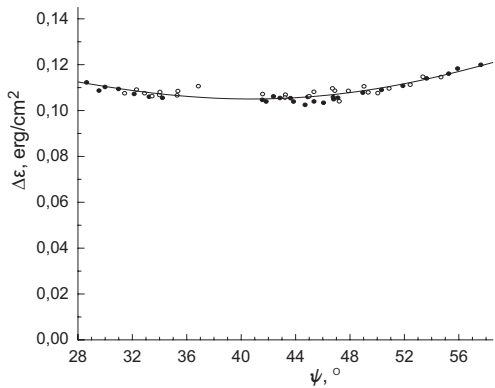}
\caption{Dependence $\Delta \varepsilon(\psi)$ plotted from the
data of the two series of measurements at ${T=0{,}9}$\,K (open
symbols) and (solid symbols) ${T=0{,}61}$\,K (solid line)
Calculation by Eq.\,(\ref{De}).} \label{De_img}
\end{figure}

Figure\,\ref{De_img} shows the $\Delta\varepsilon(\psi)$ function
calculated by Eqs. (\ref{RR}) and (\ref{RL}) using the
experimental data for the two series of measurements. Parameters
$\beta$ for each series were chosen so that the imaginary
extensions of the $\Delta\varepsilon(\psi)$ functions calculated
at ${\psi<\psi_-}$ and ${\psi>\psi_+}$, are matched in the plateau
region. The results for the $\Delta\varepsilon(\psi)$ functions
thus obtained coincide for both series within the limits of
experimental error. We take into account the symmetry of the
helium crystal, neglect the azimuthal anisotropy, assume that
$\Delta\varepsilon(\psi)$ is an analytical function  of angles,
and parametrize it as (in erg/cm$^2$)
\begin{equation}\label{De}
\Delta\varepsilon(\psi)=0{,}128-0{,}013\cos2\psi+0{,}022\cos4\psi,
\end{equation}
The calculated $\theta_0(\psi)$ dependence with allowance for the
anisotropy of $\Delta{\varepsilon}(\psi)$ (see Eq.\,(\ref{De}))
are shown as the solid lines in\,Fig.\ref{theta_psi}.

\begin{figure}[t]\center
\includegraphics[width=0.75\columnwidth]{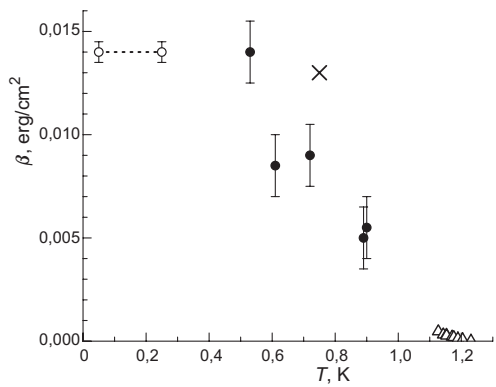}
\caption{Dependence $\beta(T)$.}\label{beta_T}
\end{figure}

Using this scheme and Eq.\,(\ref{De}), we then determined the
values of $\beta$  for the other series. Note that the neglect of
anisotropy $\varepsilon_s(\psi)$ leads to an increase in the
estimate of $\beta$ by about 30\%. The obtained temperature
dependence of $\beta$ is shown in Fig.\,\ref{beta_T} (solid
circles). Here, we also present the results from \cite{RGCB} (open
circles), which were obtained in the temperature range
50$\div$250\,K by an analysis of a spectrum of crystallization
waves, and from \cite{RGBRN} (open triangles), which were obtained
by an analysis of the (0001) face growth kinetics at temperatures
close to the roughening temperature, and the value of $\beta$
obtained in \cite{KSS} (cross) at a temperature of 0.72\,K.

Thus, our data on $\beta$ characterizing the surface energy at
angles ${0,01<\theta\lesssim0,1}$\,rad agree with the data in
\cite{RGCB} obtained at ${\theta\lesssim0,01}$\,rad. However, the
authors of \cite{RGCB} detected a sharp decrease in
$\tilde{\alpha}$ at temperatures lower than 0.25\,K and angles
${\theta\lesssim0,01}$\,rad, which agrees with the theoretical
concepts of vicinal surfaces, and  we detected non-analytical
contribution $\beta|\theta|$ to the surface energy in the angular
range ${0,01<\theta\lesssim0,1}$\,rad, which cannot be explained
theoretically at finite and weakly
temperature-dependent\,$\tilde{\alpha}.$

We thank A.F.\,Andreev, A.Ya.\,Parshin, and E.R.\,Podolyak for
fruitful discussions. This work was performed in terms of the
fundamental research program Quantum Mesoscopic and Disordered
Structures of the Presidium of the Russian Academy of Sciences.

\renewcommand{\refname}{}

Translated by K.\,Shakhlevich

JETP {\bf116}(4), 587 (2013)

\begin{thebibliography}{9}
\bibitem{KSS} K.O.\,Keshishev, V.N.\,Sorokin, and D.B.\,Shemyatikhin, JETP Lett.
{\bf85}(3), 179 (2007)
\bibitem{KS} K.O.\,Keshishev, D.B.\,Shemyatikhin, JLTP {\bf150}, 282 (2008)
\bibitem{AM} J.F.\,Allen, A.D.\,Misener, Math. Proc. of the Cambridge Phil. Soc. {\bf34}, 299 (1938)
\bibitem{RGBRN} P.E.\,Wolf, F.\,Gallet, S.\,Balibar, P.\,Nozier\`{e}s,
J.\,Physique\,(France) {\bf46}, 1987 (1985)
\bibitem{BKP} A.V.\,Babkin, D.B.\,Kopeliovich, and A.Ya.\,Parshin, Sov. Phys. JETP {\bf62}(6), 1322 (1985)
\bibitem{AKO}  O.A.\,Andreeva, K.O.\,Keshishev, and S.Yu.\,Osip'yan,
JETP Lett. {\bf49}(12), 759 (1989)
\bibitem{RGCB} E.\,Rolley, C.\,Guthmann, E.\,Chevalier,
S.\,Balibar, JLTP {\bf99}, 851 (1995)
\bibitem{MP} V.I.\,Marchenko and A.Ya.\,Parshin, JETP Lett.  {\bf83}(9), 416 (2006).
\bibitem{AK91} O.A.\,Andreeva and K.O.\,Keshishev,
Phys. Scr., {\bf39}, 325 (1991).
\end{thebibliography}
\end{document}